# Characterization of a fabrication process for the integration of superconducting qubits and RSFQ circuits

**Maria Gabriella Castellano[1], Leif Grönberg[2], Pasquale Carelli[3], Fabio Chiarello[1], Carlo Cosmelli[4], Roberto Leoni[1], Stefano Poletto[1,5], Guido Torrioli[1], Juha Hassel[2], Panu Helistö[2]**

[1] Istituto di Fotonica e Nanotecnologie, CNR, via Cineto Romano 42, I-00156 Roma, Italy
[2] VTT, P.O. Box 1000, 02044 VTT, Finland
[3] Dipartimento di Energetica, Università dell'Aquila, Monteluco di Roio, I-67040 L'Aquila, Italy
[4] Dipartimento di Fisica, Università di Roma "La Sapienza," P.le Aldo Moro 5, I-00185 Roma, Italy
[5] Dipartimento di Fisica, Università di Roma 3, via della Vasca Navale 84, I-00146 Roma, Italy

E-mail: mgcastellano@ifn.cnr.it

In order to integrate superconducting qubits with rapid-single-flux-quantum (RSFQ) control circuitry, it is necessary to develop a fabrication process that fulfills at the same time the requirements of both elements: low critical current density, very low operating temperature (tens of milliKelvin) and reduced dissipation on the qubit side; high operation frequency, large stability margins, low dissipated power on the RSFQ side. For this purpose, VTT has developed a fabrication process based on Nb trilayer technology, which allows the on-chip integration of superconducting qubits and RSFQ circuits even at very low temperature. Here we present the characterization (at 4.2 K) of the process from the point of view of the Josephson devices and show that they are suitable to build integrated superconducting qubits.

**Keywords:** Superconductivity, Quantum Computing, SQUID.

**PACS:** 85.25.Cp, 85.25.Hv, 03.67.Lx.

9 March 2006

## 1. Introduction

On the way towards a reliable and scalable quantum computer, superconducting qubits are promising candidates that are being developed worldwide [1]. To achieve the requested performance on-chip control and readout is of the utmost importance. A viable solution is rapid-single-flux-quantum (RSFQ) logic, the most advanced superconducting digital technology, which is based on overdamped Josephson junctions [2].

Ideally, the qubits must operate at very low temperature, in an environment with low dissipation to preserve the coherent behaviour, with control and readout circuit placed as close as possible, preferably integrated on-chip. These requirements are fulfilled by single-flux-quantum digital circuits, which are fast, scalable, require a very low power and share the same fabrication technology of at least some of the qubits (e.g. phase qubits, based on Josephson junctions, and flux qubits, based on Josephson interferometers).

RSFQ logic has been investigated during many years, yielding to the development of complex circuits [3], optimized however for a range of temperatures and critical current densities different from those useful for qubits. Besides, at very low temperature the tolerable dissipated power is severely limited by the low refrigerating power of dilution refrigerators close to the base temperature. As a result, building an on-chip RSFQ circuit to control and read out qubits requires a new optimization procedure, a new design of the layout and a new fabrication process.

Recently, a process satisfying these requirements has been developed by VTT. In the process, electron thermalization of the RSFQ part is improved by the use of copper cooling fins. To minimize dissipation, a target critical current density as low as 30 A/cm$^2$ is used. Here we characterize the process from the point of view of the Josephson junctions; measurements carried out at 4.2 K assess the good quality of unshunted junctions and their potential to build qubits.

## 2. RSFQ circuits and qubit requirements

RSFQ logic relies on picosecond voltage pulses generated by overdamped Josephson junctions when they switch from the superconducting to the normal state. Whenever the current through the junction is larger than the critical current $I_C$, the superconducting phase undergoes a $2\pi$ slip, which in turn produces a voltage drop across the junction. These ultra-short voltage pulses have a quantized area (integral of voltage over time) equal to a single flux quantum, $\Phi_0=h/2e\sim2$ µV/GHz: it is this magnetic flux quantum that carries the information throughout the circuit. The pulse propagation is obtained by biasing the other junctions at a current smaller than $I_C$ but such that an incoming flux quantum makes them switch to the normal state. The junctions are shunted by an external resistor $R_N$, such that the McCumber parameter is slightly smaller than unity. The speed of operation is approximately given by f~$0.3I_CR_N/\Phi_0$; power is dissipated during the pulse (P=$I\Phi_0$f) and because of the Joule effect in the bias resistors, the latter being the dominant term.

Usually, RSFQ circuits are designed to work at 4.2 K or higher temperature, with parameters optimized for speed (up to f=700 GHz) and stability. They are typically built with Nb trilayer technology, with critical current density of 1000 A/cm$^2$ or higher, and dimensions in the micron or sub-micron range. A typical RSFQ circuit can have a power dissipation ranging between 50 nW and 10 µW, mainly due to the bias resistors: this represents a small power if the circuit operates at 4.2 K, but not acceptable if the circuit must work at milliKelvin temperature, close to the qubit, where the refrigerating power of a dilution refrigerator is hardly higher than a few µW. Moreover, at such temperatures thermal conductance becomes smaller and smaller and electrons become decoupled from the phonon bath [4]; the dissipated power can turn into an overheating of the electrons, with consequent noise degradation. This issue requires a careful analysis of the thermal budget

The way out is to decrease the dissipated power by lowering the critical current density of the junctions, and add normal metal cooling fins to improve the electron thermalization. The VTT process used for the junctions of this work has been developed to accomplish all these tasks.

Ideally, with a target critical current density of $J_C=30 A/cm^2$ and a junction critical current less than 10 µA, the RSFQ circuits work with critical voltage $I_C R_N$ of only 30 µV and a clock frequency of 2 GHz, suitable for quantum computing tasks, with only 25 pW of total dissipation for a single junction. The electron thermalization can be improved by increasing the volume of the shunt resistors; this is done by connecting the resistor to a cooling fin [4], made of a thick (800 nm) copper layer. It has been calculated that the cooling fins allow, for a typical RSFQ circuit, to reach an electron temperature of only 70 mK starting from a bath temperature of 20 mK, which is sufficient to permit placing the simplest RSFQ circuits on the same chip as the qubit [5]  For more complex circuits, instead, it is necessary to adopt a hybrid solution, where qubit and RSFQ circuit are placed on two different chips, at different temperatures

As far as the qubit is concerned, the effort is to minimize all the possible sources of decoherence. Dissipation due to the electromagnetic environment has to be avoided using proper filters. Intrinsic losses stem from the dielectrics around the junction [6] and low-voltage subgap leakage: high quality Josephson junctions are needed, with a subgap I-V characteristic as close as possible to the ideal BCS curve. Even at 4.2 K, this measurement gives information on the junction quality.

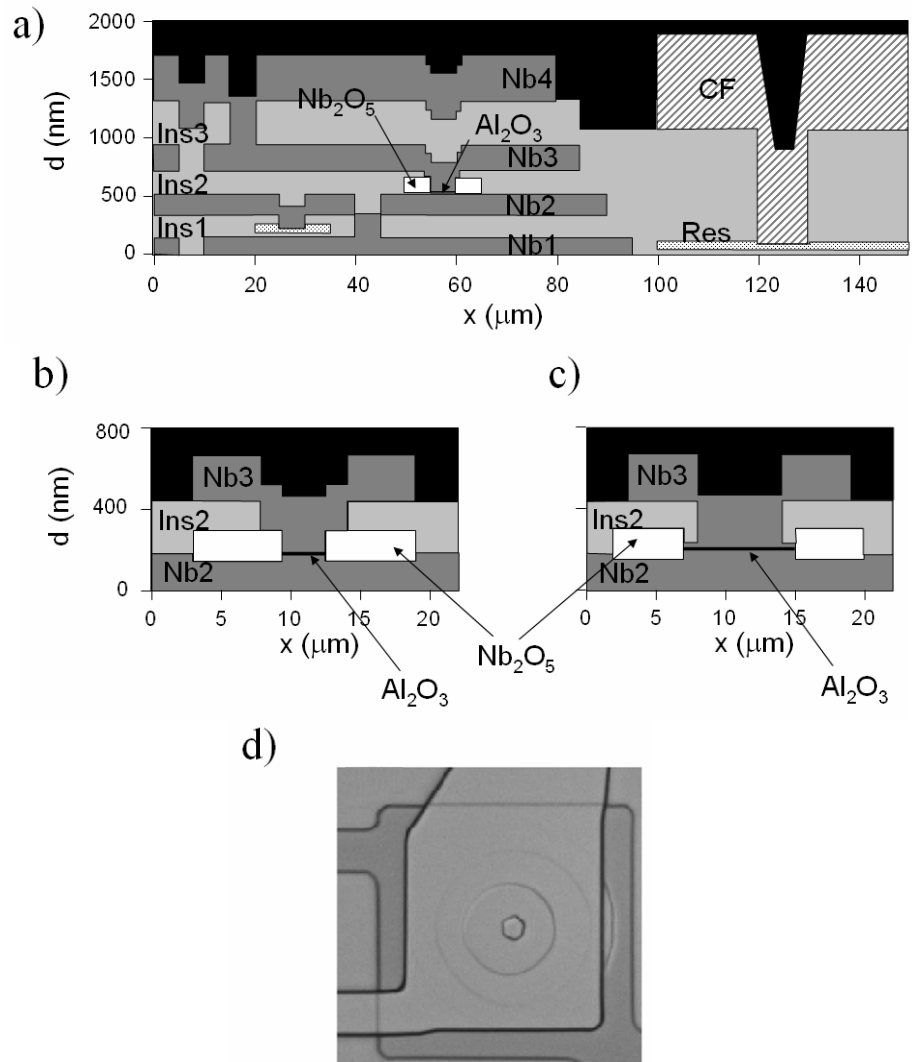

**Figure 1.** (a) Schematic stack producible by the process. Dark grey: Nb layers. Light grey: $SiO_2$ insulator. White: anodized Nb. Hatched: copper cooling fins. Dotted: Pd resistors. (b) and (c) Closeup of junctions of two different sizes. The size is defined by $Nb_2O_5$ layer. (d) A microscope photograph of a 3 µm diameter junction.

### 3. Fabrication process

The fabrication process is based on the existing VTT Nb/Al/AlO$_x$/Nb trilayer process, which has been used for about 15 years to produce SQUID magnetometers, readout amplifiers, and Josephson voltage arrays. The main modifications are increased number of superconducting layers (four), Pd resistors instead of Mo and the Cu cooling fins. A stack producible by the process is schematically shown in Fig. 1(a). Nb layers are formed by sputtered Nb. Insulator layers are SiO$_2$ formed by plasma enhanced chemical vapor deposition. Both are patterned by UV photolitography and reactive ion etching. The exception is Ins2-layer, which is patterned by wet-etching. Layers Ins1 and Ins2 are also planarized using chemical mechanical planarisation. Resistors are made of sputtered Pd and patterned by ion milling. For the Pd layer, 2nm of TiW is used to improve adhesion.

Figures 1(b) and (c) show close-ups of junction stacks and (d) shows a microscope photograph of a fabricated Josephson junction. The junctions are formed by growing Nb(200 nm)/Al(7 nm) stack, by oxidizing the surface. The oxygen exposure [7] used for the samples measured in this paper was 1000 Torr x h. After pumping the oxygen out, the Nb counter electrode is in situ grown on top. The junction geometry is then defined by anodizing the surroundings of the junction into Nb$_2$O$_5$.

### 4. Process characterization

#### 4.1. Devices

The devices tested in this work are either single, unshunted Josephson junctions of circular shape, diameter 15 µm and 6.8 µm, or low inductance dc-SQUIDs, with two circular junctions of 4.8 µm diameter. The dc-SQUID geometry is of three different types: in two of them, the SQUID hole is a double loop, arranged in a gradiometric configuration, with two different coupling schemes; in the third type, the SQUID body consists of a strip upon a ground plane. In all cases, the inductance ranges from less than 1 pH up to 6 pH and the depth of modulation of the critical current with magnetic flux is almost complete; the SQUID behaves as a single junction with twice the critical current, tunable by means of a magnetic flux. We measured at 4.2 K the I-V characteristics of 11 SQUIDs, 6 junctions with 15 µm diameter and 2 junctions with 6.8 µm diameter. An example of experimental curves is shown in fig.2. The characteristics look similar, independently on the size of the junction. The gap voltage measured over all the samples is 2.84 ± 0.03 mV. The width of the current jump at the gap, from 10% to 90%, is 92 ± 11 µV.

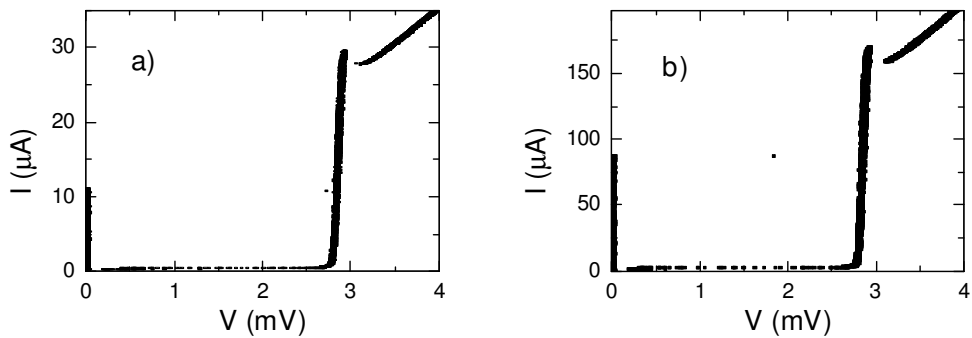

**Figure 2.** Current-voltage characteristics at 4.2 K for two Josephson devices. (a) Dc-SQUID with two circular junctions of 4.8µm diameter (b) Josephson junction, circular shape, 15 µm diameter.

#### 4.2. Critical current density

The critical current value measured by the I-V characteristics is affected by the stochastic process of thermal activation (at 4.2 K) out of the zero voltage state, resulting in underestimated values. A better estimate is obtained by analysing the statistical distribution of the switching current values [8]. To do so, the bias current is ramped until the device makes a transition from the zero-voltage state to the running state: the voltage across the device triggers the acquisition of the istantaneous bias current value. This procedure is repeated typically 10000 times. By comparing the experimental distribution of the collected data with the Kramers model [9], one gets the extrapolated critical current $I_0$ and an effective temperature for the activation process, which should be equal to the thermodynamical temperature. We performed this kind of measurement on several samples and observed that both the mean value of the switching current distribution and the extrapolated $I_0$ are a fixed fraction of the current jump at the gap, $I_g$. (Fig. 3). From our measurements, we got $I_0=0.586\ I_g$. In all the other devices, we measured the jump at the gap and inferred the value of $I_0$ using this proportionality constant.

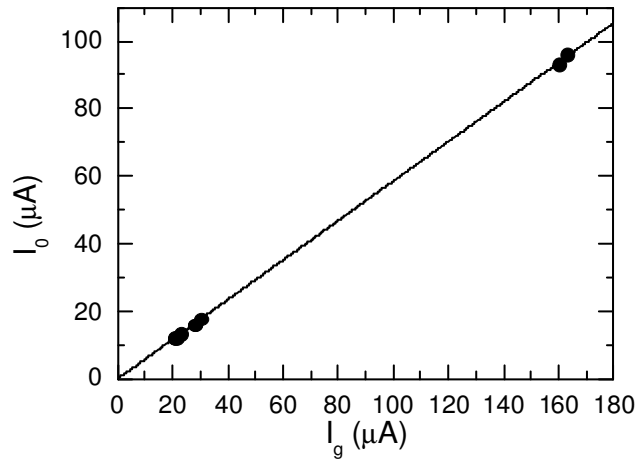

**Figure 3.** Comparison between the current $I_0$ (extrapolated from the histogram of the switching current values) and the current jump $I_g$ at the gap voltage.

To calculate the critical current density, we assume that the shrinking of the junction size in the process is negligible. This is based on a fit to room temperature resistance measurements of a large quantity of junctions with varying sizes. The best fit to the experimental data, shown in figure 4, gives $J_C=46$ A /cm$^2$.

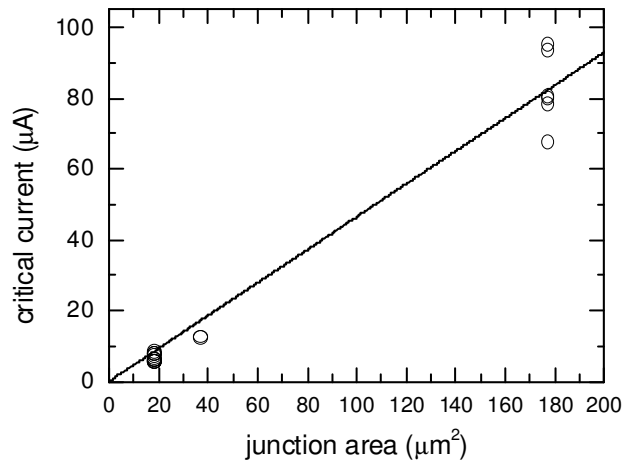

**Figure 4.** Critical current $I_0$ (inferred by current jump at the gap voltage) versus the effective area of the junctions. The line is the best fit of the data and corresponds to a critical current density of 46 A/cm$^2$.

Characterization of a fabrication process for the integration of qubits and RSFQ circuits    6

### 4.3. Junction quality

The junction quality is evaluated through the parameter $V_m=I_0R_{qp}(2\ mV)$, where $I_0$ is either measured directly with the histogram of the switching current or is calculated by the current jump at the gap voltage, and $R_{qp}(2\ mV)$ is the static resistance at 2 mV. From the histogram of the data (fig. 5), irrespective of the size of the junctions, we get $V_m=68.5\pm 5.5$ mV, a value that assesses the good quality of the junctions. The subgap residual current times the normal resistance, measured while depressing the Josephson current with a magnetic field parallel to the junction plane, is shown in fig. 6 for a junction with 15 µm diameter and a dc-SQUID (4.8 µm junctions). For comparison, the continuous line is the theoretical curve from the BCS theory, with a temperature of 4.2 K and a gap voltage of 1.425 meV: the experimental data do not depend on the size of the junctions and are only 1.2 times larger than expected for an ideal device. This is in line with other fabrication processes for superconducting qubits [10].

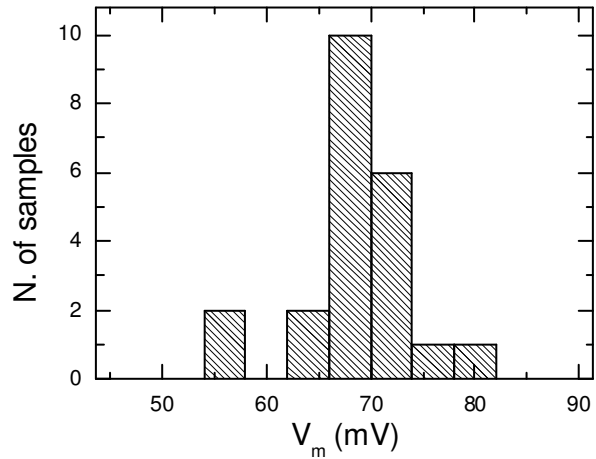

**Figure 5.** Histogram of the values of Vm.

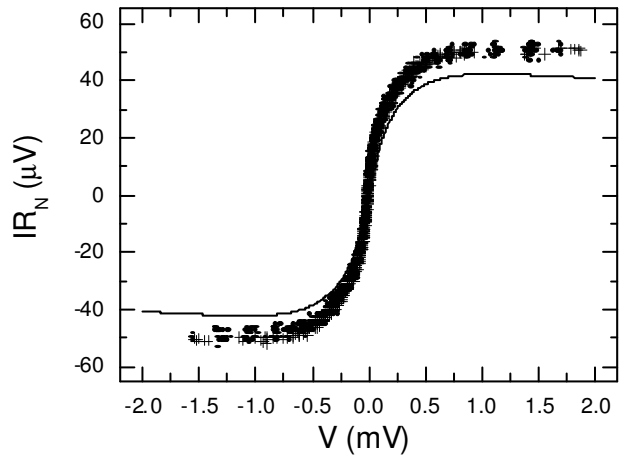

**Figure 6.** Residual subgap current times normal resistance of a Josephson junction, 15 µm diameter (crosses) and of a dc-SQUID with 4.8 µm diameter junctions (solid dots). The two experimental curves

overlap. The continuous line is the theoretical curve from the BCS theory, with a temperature of 4.2 K and a gap voltage of 1.425 meV.

### 5. Summary and Discussion

We have shown that the junction quality produced by the process is good, which encourages towards fabrication of qubit circuits. This is encouraging also from the point of view of integrated RSFQ-qubit systems. Regarding other potential sources of decoherence, we have previously measured the loss of our PECVD $SiO_2$ at 70 GHz [11], and the result can be mapped to the value of loss tangent $\tan\delta \approx 0.003$. This is of the same order as previously reported for a process known to produce functional Josephson phase qubits [6]. It is probable, though, that decoherence could be decreased by material changes, which will be considered in future. Especially the effect of $Nb_2O_5$ bounding of the junction is not known. Its contribution is likely to be significant especially in case of small junctions (see Fig. 1(b)).

This work is supported by the European Community Project RSFQubit (FP6-3749) and by the Academy of Finland (Centre of Excellence in Low Temperature Quantum Phenomena and Devices). The authors wish to thank Mikko Kiviranta, Jari S. Penttilä, Heikki Seppä and Markku Ylilammi for useful discussions.